\documentclass[%
 aip,
 twocolumn,
 amsmath,amssymb,
 reprint,%
]{revtex4-1}

\usepackage{amssymb}
\usepackage{amscd}
\usepackage{verbatim,ifthen}
\usepackage{color}
\usepackage{latexsym}

\usepackage{graphicx}
\usepackage{dcolumn}
\usepackage{bm}

\usepackage[T1]{fontenc}
\usepackage{mathptmx}

\usepackage[export]{adjustbox}

\usepackage{graphicx}
\usepackage{caption}
\usepackage{amsmath}
\usepackage{amsthm}
\usepackage{amsfonts}
\usepackage{sidecap}
\usepackage{mathtools}
\usepackage{adjustbox}
\usepackage{ upgreek }
\usepackage{}
\usepackage{soul,xcolor}

\begin{document}
\setstcolor{red}



\title[]{Machine-learned impurity level prediction for semiconductors: the example of Cd-based chalcogenides}


\author{Arun Mannodi-Kanakkithodi}
 \email{mannodiarun@anl.gov}
\affiliation{Center for Nanoscale Materials, Argonne National Laboratory, Argonne, Illinois 60439, USA}%

\author{Michael Y. Toriyama}
\affiliation{Center for Nanoscale Materials, Argonne National Laboratory, Argonne, Illinois 60439, USA}%

\author{Fatih G. Sen}
\affiliation{Center for Nanoscale Materials, Argonne National Laboratory, Argonne, Illinois 60439, USA}%

\author{Michael J. Davis}
\affiliation{Chemical Sciences and Engineering Division, Argonne National Laboratory, Argonne, Illinois 60439, USA}%

\author{Robert F. Klie}
\affiliation{Department of Physics, University of Illinois at Chicago, Chicago Illinois 60607}%

\author{Maria K.Y. Chan}%
 \email{mchan@anl.gov}
\affiliation{Center for Nanoscale Materials, Argonne National Laboratory, Argonne, Illinois 60439, USA}%

\date{\today}

\maketitle



\textbf{The ability to predict the likelihood of impurity incorporation and their electronic energy levels in semiconductors is crucial for controlling its conductivity, and thus the semiconductor's performance in solar cells, photodiodes, and optoelectronics. The difficulty and expense of experimental and computational determination of impurity levels makes a data-driven machine learning approach appropriate. In this work, we show that a density functional theory-generated dataset of impurities in Cd-based chalcogenides CdTe, CdSe, and CdS can lead to accurate and generalizable predictive models of defect properties. By converting any \textit{semiconductor + impurity} system into a set of numerical descriptors, regression models are developed for the impurity formation enthalpy and charge transition levels. These regression models can subsequently predict impurity properties in mixed anion CdX compounds (where X is a combination of Te, Se and S) fairly accurately, proving that although trained only on the end points, they are applicable to intermediate compositions. We make machine-learned predictions of the Fermi-level-dependent formation energies of hundreds of possible impurities in 5 chalcogenide compounds, and we suggest a list of impurities which can shift the equilibrium Fermi level in the semiconductor as determined by the dominant intrinsic defects. These `dominating' impurities as predicted by machine learning compare well with DFT predictions, revealing the power of machine-learned models in the quick screening of impurities likely to affect the optoelectronic behavior of semiconductors.}





\section*{Introduction}

No crystalline material is devoid of defects and impurities. In fact, the imperfections in a crystal determine its properties as much as the regular arrangement of atoms do. When it comes to crystalline semiconducting materials, it is known that defects such as vacancies, native or impurity interstitials or substitutions, surface states, and grain boundaries can influence their optoelectronic properties. In the absence of external impurities, native defects determine the equilibrium Fermi level in the semiconductor, and thus the nature of conductivity (p-type, n-type or intrinsic) and charge carriers \cite{Def1,Def2,Def3}. The introduction of impurity atoms can change the conductivity as determined by the dominant native defects, based on their formation enthalpies as a function of the Fermi energy \cite{Def1,fermi_eqm}. Foresight about the impact of certain impurities on the electronic structure and conductivity of the material is crucial in either trying to curb their presence, or intentionally incorporating them in the semiconductor lattice to induce a desirable optoelectronic change. \\

It is important to be able to predict the electronic energy levels created by impurities in semiconductors. While shallow acceptor or donor levels are defined as defect levels close to the band edges and do not affect the recombination of charge carriers, deep defect levels can have both disastrous and potentially beneficial effects. Deep levels can act as non-radiative recombination centers for minority charge carriers, which significantly reduces their lifetime, impedes carrier collection or light emission, and drastically brings down the solar cell or photodiode efficiency and performance \cite{Defect_levels_4}. On the other hand, researchers have shown that in principle, energy levels in the band gap can be used as \textit{intermediate bands} to facilitate absorption of sub-gap photons, which could enhance the absorption efficiencies \cite{Def1,IB1,IB2}. \\

\begin{figure*}[t]
\includegraphics[width=\linewidth]{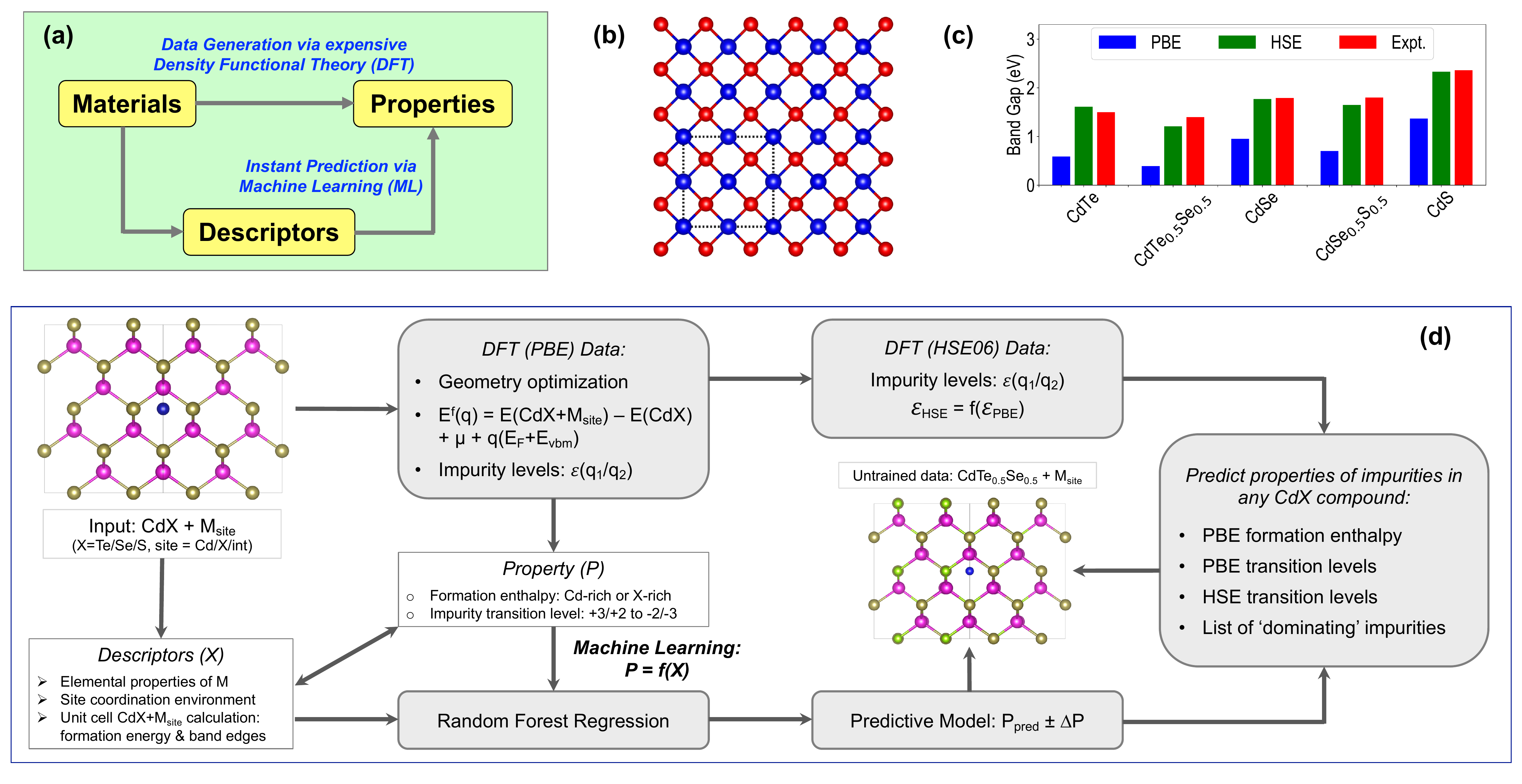}
\caption{\label{Fig:outline} 
(a) General outline of materials design process leading to ML-driven prediction of properties based on DFT data and intermediate step of converting materials to numerical descriptors. (b) The Zinc Blende structure adopted by CdTe, CdSe, and CdS. Cd atoms are shown in blue and Te/Se/S atoms in red. The unit cell has been indicated with dashed lines. (c) Comparison of band gaps computed at the PBE and HSE06 levels of theory with reported experimental values \cite{Expt_gap1,Expt_gap2}, for CdTe, CdSe, CdS, CdTe$_{0.5}$Se$_{0.5}$ and CdSe$_{0.5}$S$_{0.5}$. (d) Outline of the DFT and ML driven prediction of properties of impurities in Cd-based chalcogenides.}
\end{figure*}

Defect levels are often measured using methods like deep level transient spectroscopy (DLTS) and cathodoluminescence (CL) \cite{DefectExpt1,DefectExpt2,DefectExpt3,DefectExpt4}. However, difficulties in incorporating specific impurities or dopants in a given compound and in attributing measured levels to specific defects make experimental methods less than ideal for an extensive study of defects and impurities in semiconductors. First-principles density functional theory (DFT) computations have been widely used instead to simulate substitutional or interstitial impurities and vacancies in crystalline materials using the supercell approach \cite{Defect_DFT_1,Defect_DFT_2,Corr2}. Impurity formation enthalpies, energy levels, and resulting absorption coefficients calculated from DFT typically match well with measured values \cite{DFT_expt_1,DFT_expt_2,DFT_expt_3,DFT_expt_4,DFT_expt_5}. However, DFT has limitations of its own: the requirement of large supercells, charge states, explicit charge image corrections, and an advanced level of theory (such as hybrid functionals \cite{HSE_gap} or $GW$ corrections \cite{GW}) to accurately determine band gaps make these calculations generally expensive. Furthermore, prior knowledge is seldom utilized in informing or accelerating new defect calculations; there is an opportunity here for the creation of surrogate models based on previously generated data, such that impurity properties for fresh cases can be quickly and accurately estimated. \\

Today, machine learning (ML) has become an integral component of materials design \cite{ML1}. Researchers have extracted models and design rules from materials data to drive the accelerated discovery of NiTi alloys for thermal hysteresis \cite{ML6}, design of polymer dielectrics for improved energy storage in capacitors \cite{ML7,ML8}, synthesis of new classes of compounds \cite{ML9,ML10}, identification of new and improved catalysts \cite{ML11,ML12}, and the design of experiments in a smart and `adaptive' fashion \cite{ML13}. ML-based design of materials usually begins with the generation of sufficient data for candidate materials in terms of a property \textit{\textbf{P}}, and the conversion of all materials in the chemical space into a unique numerical representation \textit{\textbf{X}}, referred to as descriptors, feature vectors, or fingerprints. This is followed by a mapping \textit{\textbf{X}} $\rightarrow$ \textit{\textbf{P}} between descriptors and properties using linear correlation \cite{Def1,ML15} or a nonlinear regression technique such as ridge regression \cite{ML16}, support vector machine  \cite{ML17}, random forest \cite{ML18}, LASSO (Least Absolute Shrinkage and Selection Operator) \cite{ML19}, or neural networks \cite{ML20}. The result of such an approach is a trained predictive model which estimates \textit{\textbf{P}} for any \textit{\textbf{X}}, with a statistical uncertainty or confidence interval that is also an output. The general outline for developing machine-learned predictive models for properties of materials based on DFT data and numerical descriptors is shown in Fig. \ref{Fig:outline}(a). \\

The prediction of defect or impurity formation enthalpies and energy levels can be accelerated by developing ML models trained from DFT data, as has been shown in the recent past \cite{ML_Varley}. As a demonstration of this approach, we take the example of Cd-based chalcogenides, which are important semiconductors for optoelectronic and solar cell applications \cite{CdTe1,CdTe2,CdTe3}, and apply ML algorithms on a dataset of DFT computed properties for hundreds of impurity types in CdTe, CdSe, and CdS. These compounds are chosen not only because CdTe-based cells are the second most commonly used photovoltaics after Si, but also because in recent years, significant improvements in the efficiency of CdTe solar cells have arisen due to the elimination of the CdS buffer layer and the introduction of Cd(Se,Te) into the absorber layer\cite{CdSeTe}. Therefore, the prediction of impurity levels in ternary Cd chalcogenides of various compositions is of technological importance.  Each of these compounds, henceforth referred to as CdX (X = Te/Se/S), exists in the cubic Zinc Blende (ZB) structure \cite{ZB} shown in Fig. \ref{Fig:outline}(b). The band gaps are underestimated by the PBE functional \cite{PBE_gap} but match well with measured values when using the HSE06 hybrid functional \cite{HSE_gap}; while this effect may translate to impurity and defect levels as well, it has been shown in the past that the PBE functional, despite its well known band gap underestimation, can sometimes capture the defect transition levels to span the physical band gap \cite{Defect_DFT_6}. In Fig. \ref{Fig:outline}(c), we plotted the band gaps computed from PBE and HSE06 functionals alongside the known, experimentally measured band gaps \cite{Expt_gap1,Expt_gap2} for 5 compounds: CdTe, CdSe, CdS, and mixed anion compounds (simulated using anion-ordered structures), CdTe$_{0.5}$Se$_{0.5}$ and CdSe$_{0.5}$S$_{0.5}$; the HSE06 computed values match well with experiments. The DFT computed lattice constants and band gaps for the 5 compounds are listed in Table SI-1. \\

\begin{figure*}[t]
\includegraphics[width=\linewidth]{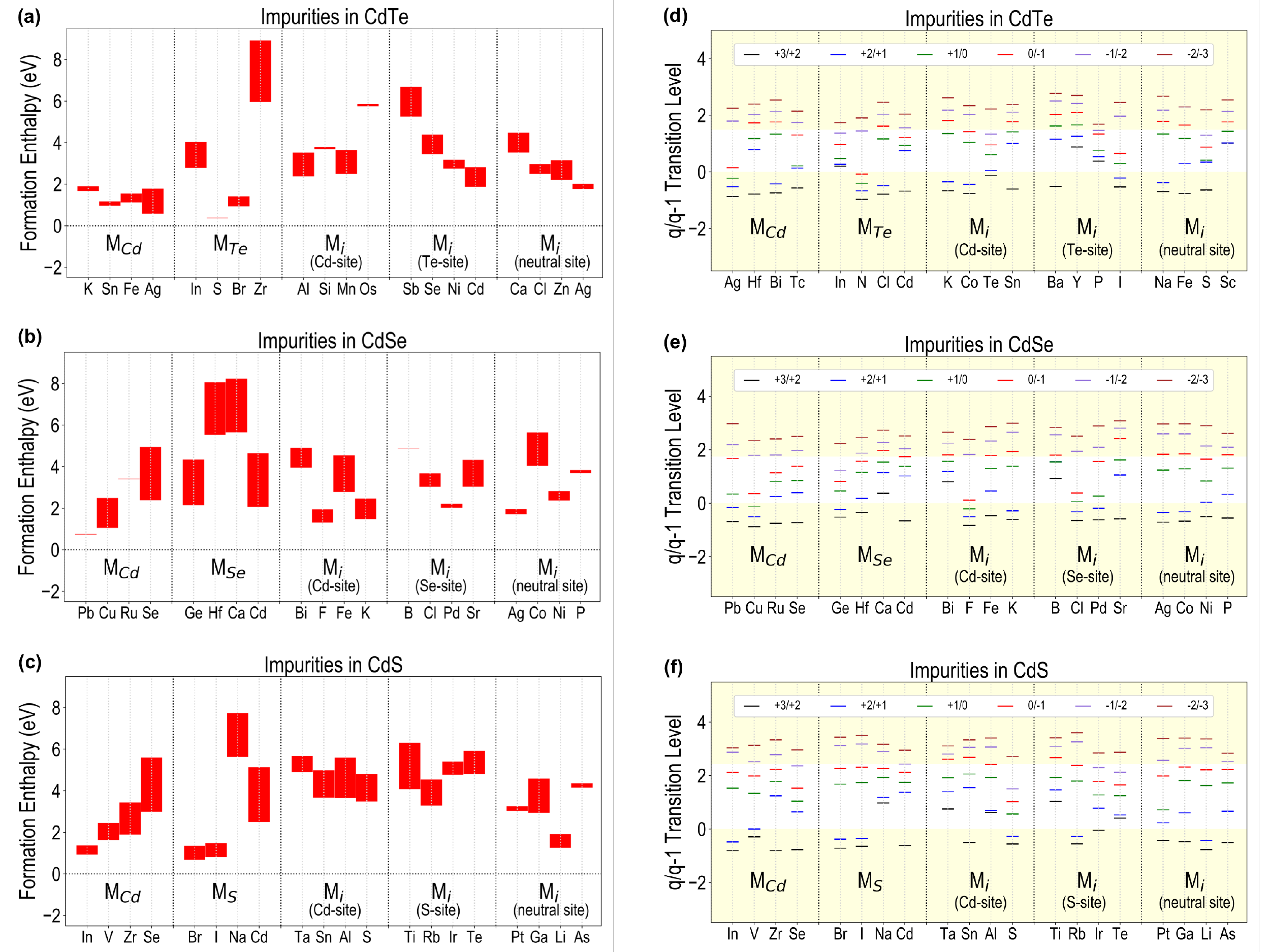}
\caption{\label{Fig:PBE_data} Neutral-state impurity formation enthalpies computed at the PBE level of theory for selected impurity atoms in different sites in (a) CdTe, (b) CdSe and (c) CdS, and charge transition levels (from +3/+2 to -2/-3) calculated at the PBE level of theory for selected impurity atoms in different sites in (d) CdTe, (e) CdSe and (f) CdS. $\Updelta$H has been plotted for some very unstable impurities as well (like Hf$_\mathrm{Se}$ and Ta$_\mathrm{i}$) to show the variety in the impurity property data that goes into training predictive models.}
\end{figure*}

In this work, we use both the PBE and HSE06 functionals to compute impurity properties in different CdX compounds; the eventual dataset of HSE impurity levels is one-fifth the size of the corresponding PBE dataset, owing to the 2 orders of magnitude difference in computational expense. We train separate ML models for impurity properties computed with PBE and HSE, and explore how models trained for lower fidelity (presumably, PBE) can inform the higher fidelity (presumably, HSE) predictions. We simulate impurities in several different defect sites in any CdX compound: one cation site (M$_\mathrm{Cd}$, where M is the impurity atom), one or two anion sites (M$_\mathrm{X}$) and three or four interstitial sites (M$_\mathrm{i}$), based on whether it is a pure or mixed anion composition \cite{Dmitry}; each of these sites have been pictured for CdTe in Fig. SI-1. Impurity atoms M are obtained by sweeping across the periodic table and selecting elements from periods II to VI, as shown in Fig. SI-2. \\

An outline of the work presented in this manuscript is shown in in Fig. \ref{Fig:outline}(d). DFT is used to compute the impurity formation enthalpy as a function of chemical potential ($\mu$), charge ($q$) and Fermi energy (E$_{F}$), using equation \ref{eqn-1}, and the impurity charge transition levels using equation \ref{eqn-2}. ML models are trained for two types of properties: the neutral-state formation enthalpy $\Updelta$H (E$^{f}$($q$=0) for Cd-rich to X-rich chemical potential conditions), and various impurity charge transition levels, $\epsilon$(q$_1$/q$_2$), which indicates the Fermi level at which the impurity containing system transitions from one stable charge state (q$_1$) to another (q$_2$). As shown in Fig. \ref{Fig:outline}, descriptors are generated for any CdX + M$_\mathrm{site}$ system (where M and `site' refer to the impurity atom and defect site respectively) based on tabulated elemental properties of M (such as ionic radii and electronegativity), site coordination environment, and properties computed from low-cost unit cell defect calculations. A regression algorithm is applied to map the descriptors to the properties, and predictive models are trained on the PBE formation enthalpy, PBE impurity transition levels, and HSE impurity transition levels. Comparisons in ML performance are made for different sets of descriptors, level of theory (PBE or HSE), and subset of computational data used for training. While models are trained for impurities in CdTe, CdSe and CdS, we performed additional computations for selected impurities in CdTe$_{0.5}$Se$_{0.5}$ and CdSe$_{0.5}$S$_{0.5}$, to test the models' \textit{out-of-sample} predictive ability. The power of this combined DFT + ML approach is illustrated with machine-learned predictions of Fermi level dependent formation enthalpies for the entire chemical space of impurities in CdTe, CdSe, CdS, CdTe$_{0.5}$Se$_{0.5}$, and CdSe$_{0.5}$S$_{0.5}$. These predictions, combined with the DFT computed formation enthalpies of intrinsic point defects (vacancies, anti-site and interstitials) in each of the compounds, are used to obtain the list of impurities which can shift the equilibrium Fermi level (as determined by dominant native defects) and thus change the nature of conductivity in the semiconductor. \\

\section*{RESULTS AND DISCUSSION}

\begin{table*}[ht]
\centering
\caption{Details of the DFT dataset.}
\label{table:dataset}
\medskip
\begin{tabular}{c|c|c|c|c|c|c|c}
\hline
\textbf{Property} & \textbf{CdX} & \textbf{Impurity Atoms} & \textbf{Defect Sites} & \textbf{Transition Levels} & \textbf{Total Chemical Space} & \textbf{DFT Data} & \textbf{$\%$ of Computed Data} \\
\hline
                 &            &        &    &    &                                       &          &        \\
                 &  CdTe & 63  & 5 & - &                  63*5 = 315  &  315  &  100  \\
                 &  CdSe & 63  & 5 & - &                  63*5 = 315  &  315  &  100  \\
PBE $\Updelta$H  &  CdS &  63  & 5 & - &                  63*5 = 315  &  315  &  100  \\
                 &  CdTe$_{0.5}$Se$_{0.5}$ & 63 & 7 & - &  63*7 = 441  &  22  & $\sim$ 5   \\
                 &  CdSe$_{0.5}$S$_{0.5}$  & 63 & 7 & - &  63*7 = 441  &  22  & $\sim$ 5   \\
\hline
                 &  \textbf{Total}  &  &  &                  &  1827        &  989   & $\sim$ 54  \\
\hline
                 &            &        &    &    &                                       &          &        \\
                             &  CdTe &  63  & 5 & 6 &                   63*5*6 = 1890  &  1890 &  100   \\
                             &  CdSe &    63  & 5 & 6 &                 63*5*6 = 1890  &  198  &  $\sim$ 10.5  \\
PBE $\epsilon$(q$_1$/q$_2$)  &  CdS &    63  & 5 & 6 &                  63*5*6 = 1890  &  198  &  $\sim$ 10.5  \\
                             &  CdTe$_{0.5}$Se$_{0.5}$ &  63  & 7 & 6 & 63*7*6 = 2646  &  132  &  $\sim$ 7   \\
                             &  CdSe$_{0.5}$S$_{0.5}$  &  63  & 7 & 6 & 63*7*6 = 2646  &  132  &  $\sim$ 7   \\
\hline
                 &  \textbf{Total}    &  &  &                &  10962        &  2550   &   $\sim$ 23   \\
\hline
                 &            &        &    &    &                                       &          &        \\
                             &  CdTe &  63  & 5 & 4 &                    63*5*4 = 1260  &  240  &  $\sim$ 19  \\
                             &  CdSe &   63  & 5 & 4 &                   63*5*4 = 1260  &  132  &  $\sim$ 10.5  \\
HSE $\epsilon$(q$_1$/q$_2$)  &  CdS &  63  & 5 & 4 &                     63*5*4 = 1260  &  132  &  $\sim$ 10.5   \\
                             &  CdTe$_{0.5}$Se$_{0.5}$ & 63  & 7 & 4 &   63*7*4 = 1764  &  88   &  $\sim$ 5   \\
                             &  CdSe$_{0.5}$S$_{0.5}$  & 63  & 7 & 4 &   63*7*4 = 1764  &  88   &  $\sim$ 5   \\
\hline
                 &  \textbf{Total}  &  &  &                 &  7308        &  680   &  $\sim$ 9.3  \\
\hline
\end{tabular}
\end{table*}

\subsection{PBE Data: Formation Enthalpy and Transition Levels}

The zero charge version of Equation \ref{eqn-1} was used to compute the formation enthalpy $\Updelta$H of impurities in CdX at Cd-rich and X-rich chemical potential conditions, for a few hundred impurity types. For CdTe, CdSe, and CdS, the neutral state impurity calculations are performed for each of the 63 elemental impurities as shown in Fig. SI-2, leading to a dataset of 315 $\Updelta$H ranges (Cd-rich to X-rich) for each compound. The computed $\Updelta$H ranges have been plotted for the entire dataset in Figures SI-3, SI-4 and SI-5, and for a few selected cases in Fig. \ref{Fig:PBE_data}. It can be seen from Fig. \ref{Fig:PBE_data} (a), (b), and (c) that anti-site substitutional impurities such as Zr$_\mathrm{Te}$, Se$_\mathrm{Cd}$, and Na$_\mathrm{S}$ have high formation enthalpies and would be unstable, whereas other impurities like Ag$_\mathrm{Cd}$, S$_\mathrm{Te}$, and Br$_\mathrm{S}$ have much lower formation enthalpies. Further, 22 impurities were selected in CdTe$_{0.5}$Se$_{0.5}$ and CdSe$_{0.5}$S$_{0.5}$ across all 7 defect sites, and $\Updelta$H was computed for each to test the trained ML models (explained in the coming sections). The formation enthalpies are shown in Fig. SI-6 and Fig. SI-7. A description of the PBE $\Updelta$H dataset across the 5 CdX compounds is provided in Table \ref{table:dataset}; data was generated for over 50$\%$ of the total chemical space of 1827 points. \\

Next, supercells containing impurity atoms were simulated in charge states of +3, +2, +1, -1, -2 and -3. For each of these calculations, the total DFT energies and charge correction terms (using Freysoldt's correction \cite{Corr1,Corr2}) were obtained, and equation \ref{eqn-2} was used to compute the various charge transition levels. All computed transition levels, namely, +3/+2, +2/+1, +1/0, 0/-1, -1/-2 and -2/-3, are plotted for the entire dataset of impurities in different sites in CdTe, CdSe, and CdS in Figures SI-8, SI-9, and SI-10, respectively. This data has been presented once again for select impurities in Fig. \ref{Fig:PBE_data} (d), (e) and (f). It should be noted that on occasion, transition levels like +1/-1 or +2/0 may exist, in which case the q/(q-1) and (q-1)/(q-2) transition levels are considered to be equal to the q/(q-2) transition level (for eg., +1/0 = 0/-1 = +1/-1). From Fig. \ref{Fig:PBE_data}, it can be seen that a number of impurities induce energy levels in the band gap, which is attributed to the specific element's stability in a charge state other than +2 at the Cd-site (for instance, Bi$_\mathrm{Cd}$ displays a +1/0 transition in CdTe), -2 at the Te-site, or 0 at the interstitial site. Impurities that create mid-gap energy levels will be of interest if their formation enthalpies are low enough for them to be competitive with respect to dominant intrinsic point defects. Further, for the 22 additional impurities in CdTe$_{0.5}$Se$_{0.5}$ and CdSe$_{0.5}$S$_{0.5}$, all the transition levels are computed and plotted in Figs SI-11 and SI-12. As listed in Table \ref{table:dataset}, DFT data was generated for 100$\%$ of the CdTe points, but 10$\%$ or less for CdSe, CdS, CdTe$_{0.5}$Se$_{0.5}$, and CdSe$_{0.5}$S$_{0.5}$. The total DFT dataset covers about 23$\%$ of the chemical space, providing a great opportunity for machine learning the remaining points. \\

\subsection{Descriptors for Machine Learning}

As shown in Fig. \ref{Fig:outline}(a), the training of prediction models for material properties proceeds via the crucial intermediate step of descriptor generation. In this work, we utilize different sets of descriptors that represent the impurity atom and the defect site coordination, as well as some properties estimated from low-cost unit cell calculations. Similar descriptors were recently applied by us to represent impurities at the Pb-site in methylammonium lead bromide \cite{Def1}, from which we were able to train simple models to describe the formation enthalpy and charge transition levels. In a similar vein, we use the elemental properties of the impurity atom M, the number of Cd or X (Te/Se/S) neighboring atoms at the given defect site, and energetic and electronic properties calculated by modeling the M$_\mathrm{Cd}$, M$_\mathrm{X}$ or M$_\mathrm{i}$ impurity in an 8-atom (Zinc Blende) CdX unit cell instead of a 64-atom supercell. The unit cell calculation is two orders of magnitude cheaper than the corresponding supercell calculation. \\

We apply different combinations of descriptors and use different regression algorithms to train predictive models for $\Updelta$H and $\epsilon$(q$_1$/q$_2$). A base set of descriptors, namely the period and group of M, a defect site index (set as 0 for M$_\mathrm{Cd}$, 1 for M$_\mathrm{X}$, 0.50 for M$_\mathrm{i}$ \textit{(neutral site)}, 0.25 for M$_\mathrm{i}$ \textit{(Cd-site)} and 0.75 for M$_\mathrm{i}$ \textit{(X-site)}), and the number of Cd and X neighbors, is used in every combination. In addition, the elemental properties of M, such as the first ionization energy, electronegativity, and ionic radii, are used as descriptors to encode information about the structural and bonding characteristics of the impurity atom. Lastly, the impurity formation enthalpy at Cd-rich, intermediate, and X-rich chemical potential conditions, and the valence band and conduction band edges (universally aligned using the deep 5$s$ semi-core state of Cd) calculated from the unit cell defect calculation are added as descriptors. Ultimately, we apply the following three sets of descriptors (in addition to the base set descriptors) independently to train the models:

\begin{enumerate}

\item Elemental properties

\item Unit cell defect properties

\item Elemental properties + unit cell defect properties

\end{enumerate}

\begin{figure}[t]
\includegraphics[width=\linewidth]{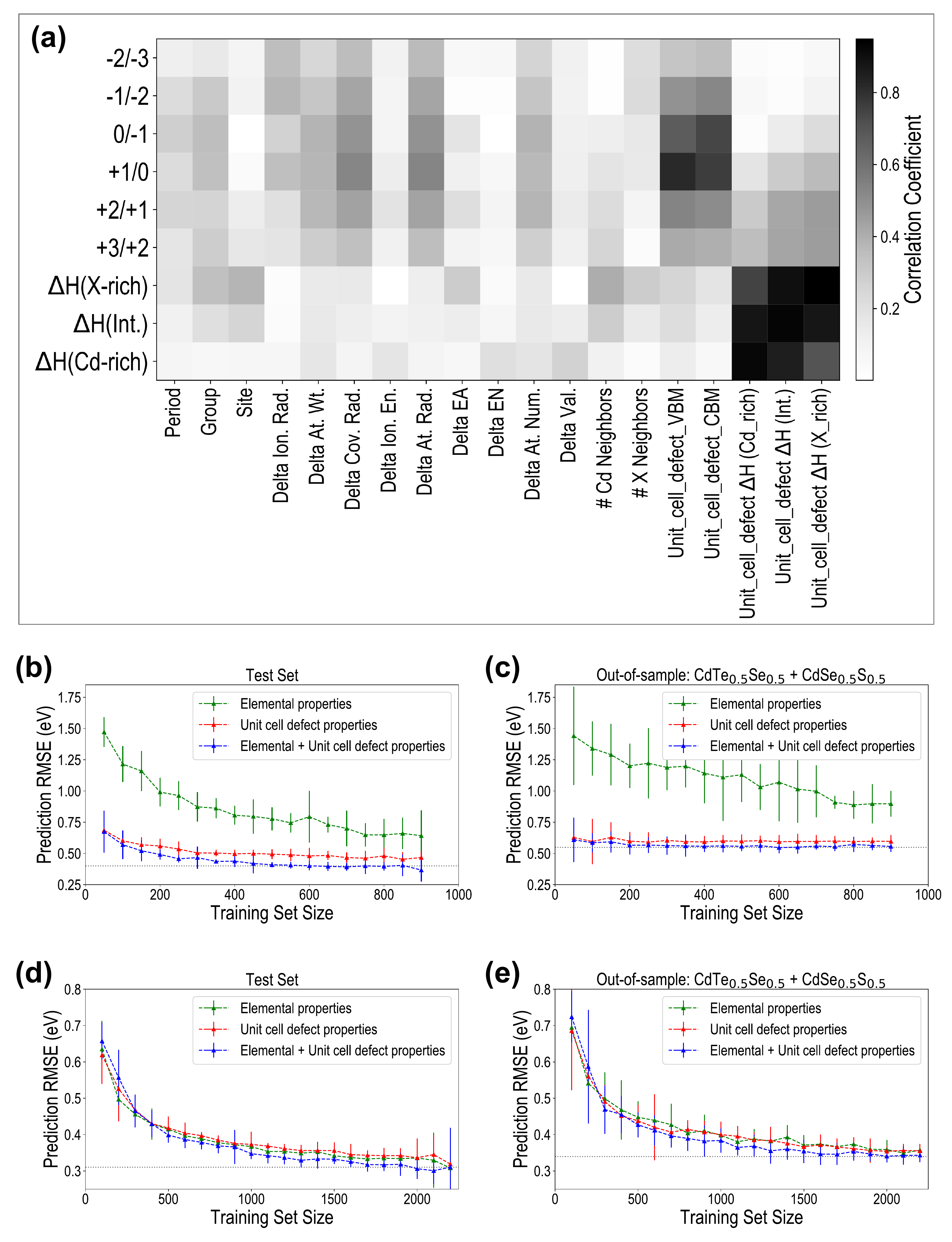}
\caption{\label{Fig:Learning} (a) Coefficient of linear correlation ($|r|$) between the properties of interest, $\Updelta$H and $\epsilon$(q$_1$/q$_2$), and each of the descriptors. In (b) and (c), prediction RMSE is plotted against the training set size for random forest models trained for $\Updelta$H (Cd-rich) using 3 different sets of features, for the test set points (total CdTe+CdSe+CdS dataset minus the training set) and the out-of-sample points (set of 22 impurities each in CdTe$_{0.5}$Se$_{0.5}$ and CdSe$_{0.5}$S$_{0.5}$) respectively. Similar plots are shown for $\epsilon$(q$_1$/q$_2$) (at the PBE level of theory) (d) test set points and (e) out-of-sample points.}
\end{figure}

In Fig. \ref{Fig:Learning}(a), we plot the Pearson correlation coefficient ($|r|$) between each descriptor and 9 different properties, namely the $\Updelta$H for Cd-rich, intermediate and X-rich conditions, and the +3/+2, +2/+1, +1/0, 0/-1, -1/-2 and -2/-3 impurity transition levels. It can be seen that while some of the elemental properties have a correlation of 0.40 to 0.50 with $\Updelta$H and $\epsilon$(q$_1$/q$_2$), the unit cell defect properties exhibit the highest correlations. The valence and conduction band edges from unit cell defect calculations show a correlation of $|r|$ = 0.82 and $|r|$ = 0.74 respectively with the +1/0 and 0/-1 impurity transition levels. Further, $\Updelta$H (Cd-rich), $\Updelta$H (intermediate) and $\Updelta$H (X-rich) show a correlation of $|r|$ > 0.90 with the corresponding $\Updelta$H  values from unit cell defect calculations. When training predictive models for the impurity formation enthalpies and transition levels using these descriptors, one can expect more accurate predictions when including the unit cell defect properties as opposed to using elemental properties exclusively. However, while the unit cell defect calculations are not computationally intensive, the remaining descriptors can be generated with no additional computations at all, and thus have an advantage. In the next section, we examine the accuracy of regression models trained using different sets of descriptors. \\

\subsection{Predictive Models using Regression}

Three regression algorithms, namely Random Forest regression (RFR) \cite{ML18}, Kernel Ridge Regression (KRR) \cite{ML16}, and LASSO regression \cite{ML19} (see details in \textit{Methods}) were applied to train predictive models for $\Updelta$H and the $\epsilon$(q/q-1) transition level for a given charge q. For each property, we trained models using the CdTe, CdSe, and CdS data, and data generated for CdTe$_{0.5}$Se$_{0.5}$ and CdSe$_{0.5}$S$_{0.5}$ was used to test the \textit{out-of-sample} predictive power. For the impurity formation enthalpy, we train separate models for $\Updelta$H (Cd-rich) and $\Updelta$H (X-rich), since the two values provide the range of possible enthalpies over the chemical potential region of stability. The effects of training set size and choice of descriptors are studied by estimating the mean and standard deviation in prediction error over 100 different models  trained (from different training sets) for any given case. \\

\begin{figure}[b]
\includegraphics[width=\linewidth]{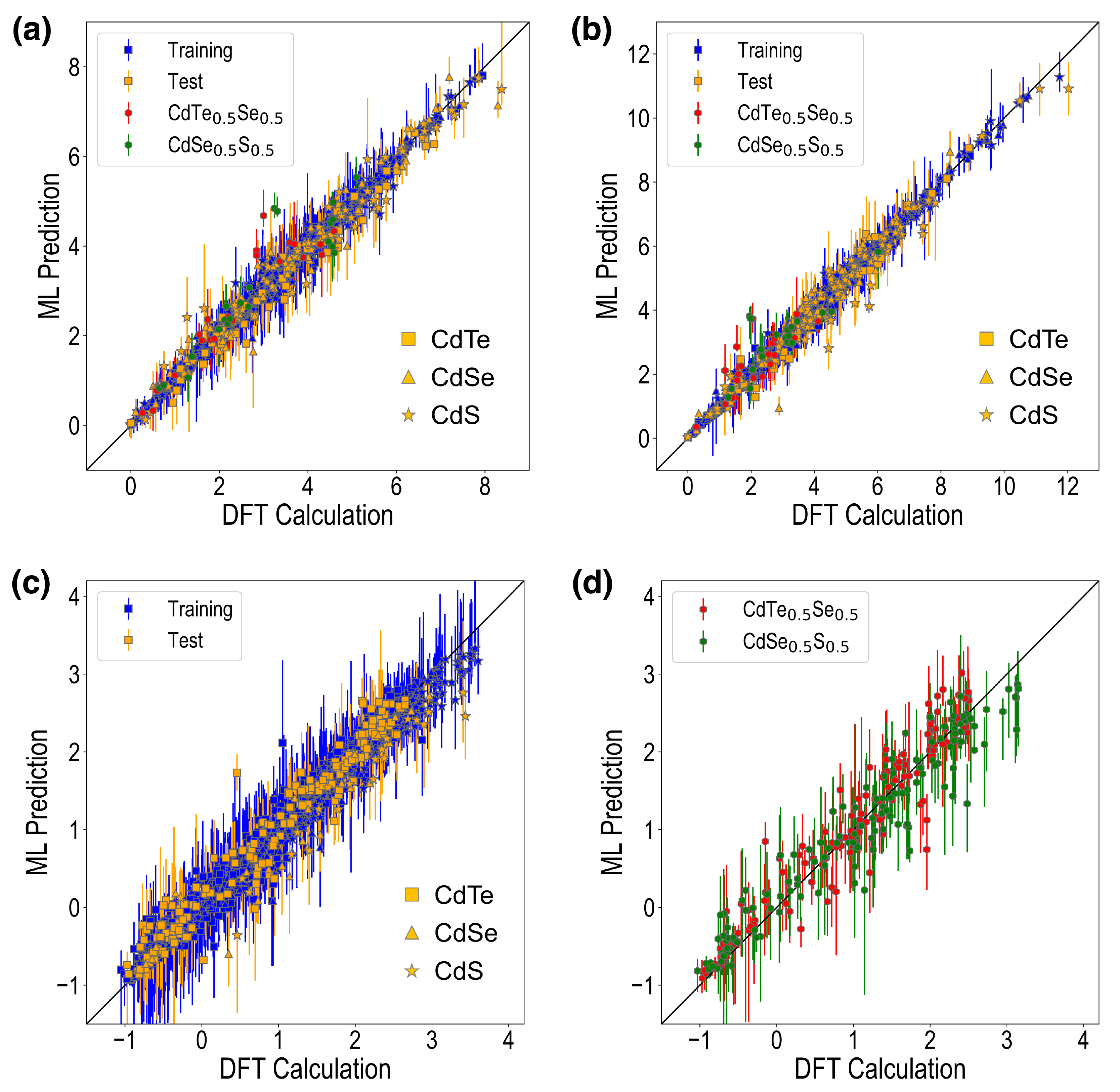}
\caption{\label{Fig:PBE_parity} Parity plots for random forest regression models trained for (a) $\Updelta$H (Cd-rich), and (b) $\Updelta$H (X-rich). Pictured are the training and test set points (the training set size is 90$\%$ of the dataset of CdTe, CdSe and CdS points), and the CdTe$_{0.5}$Se$_{0.5}$ and CdSe$_{0.5}$S$_{0.5}$ points. Similarly, parity plots are shown for models trained for $\epsilon$(q$_1$/q$_2$) (at the PBE level of theory) using the dataset of 2286 points (total CdTe+CdSe+CdS dataset), for (c) the training and test set points, and (d) the CdTe$_{0.5}$Se$_{0.5}$ and CdSe$_{0.5}$S$_{0.5}$ points.}
\end{figure}

The root mean square errors (RMSE) of RFR models trained for $\Updelta$H (Cd-rich) are plotted as a function of the training set size for three sets of descriptors (each containing the base set), for the test set in Fig. \ref{Fig:Learning}(b), and the out-of-sample points in Fig. \ref{Fig:Learning}(c). All prediction errors steadily decrease with increasing training set size. It can be seen that for both the test and out-of-sample points, using just the elemental properties as descriptors leads to much higher errors than using the unit cell defect properties. The combination of elemental and unit cell defect properties shows the best prediction accuracies, and saturate fairly early to about 0.40 eV for the test set and 0.55 eV for the out-of-sample points, proving that reasonable prediction accuracies can be achieved with about 50$\%$ of the total data used for training. In Table \ref{table:PBE_form}, we have listed the RMSE for predictive models trained using RFR, KRR, and LASSO, with 90$\%$ of the dataset of CdTe, CdSe, and CdS points used for training, independently applying the three sets of descriptors. RFR shows better performances than LASSO for every set; KRR shows slightly better test set errors than RFR, but the RFR predictions on the out-of-sample CdTe$_{0.5}$Se$_{0.5}$ and CdSe$_{0.5}$S$_{0.5}$ points are undeniably better, which gives us confidence to use the random forest models going forward. Fig. \ref{Fig:PBE_parity} shows parity plots for the best RFR models trained using 90$\%$ of data for training, for $\Updelta$H (Cd-rich) in (a) and $\Updelta$H (X-rich) in (b). The corresponding models trained using KRR and LASSO are shown for comparison in Fig. SI-14. \\

\begin{table*}[htp]
\centering
\caption{RMSE (in eV) for regression models trained for PBE $\Updelta$H (Cd-rich), using different methods and sets of features.}
\label{table:PBE_form}
\medskip
\begin{tabular}{c|c|c|c|c}
\hline
\textbf{Dataset} & \textbf{Regression Method} & \textbf{Elemental properties} & \textbf{Unit cell defect properties} & \textbf{Elemental + Unit cell defect properties}  \\
\hline
          &  RFR     &  0.40  &  0.20  &  0.17  \\
Training  &  KRR    &  0.40  &  0.30  &  0.20  \\
          &  LASSO  &  0.62  &  0.50  &  0.44  \\
\hline
      &  RFR     &  0.65  &  0.45  &  0.38  \\
Test  &  KRR    &  0.68  &  0.40  &  0.32  \\
      &  LASSO  &  0.75  &  0.52  &  0.47  \\
\hline
                        &  RFR     &  0.84  &  0.57  &  0.52  \\
CdTe$_{0.5}$Se$_{0.5}$  &  KRR    &  0.80  &  0.65  &  0.57  \\
                        &  LASSO  &  0.95  &  0.73  &  0.65  \\
\hline
                       &  RFR     &  0.86  &  0.63  &  0.57  \\
CdSe$_{0.5}$S$_{0.5}$  &  KRR    &  0.75  &  0.68  &  0.70  \\
                       &  LASSO  &  0.92  &  0.70  &  0.72  \\
\hline
\end{tabular}
\end{table*}

\begin{table*}[htp]
\centering
\caption{RMSE (in eV) for regression models trained for PBE $\epsilon$(q$_1$/q$_2$), using different methods and sets of features.}
\label{table:PBE_CT}
\medskip
\begin{tabular}{c|c|c|c|c}
\hline
\textbf{Dataset} & \textbf{Regression Method} & \textbf{Elemental properties} & \textbf{Unit cell defect properties} & \textbf{Elemental + Unit cell defect properties}  \\
\hline
          &  RFR     &  0.18  &  0.15  &  0.13  \\
Training  &  KRR    &  0.27  &  0.28  &  0.25  \\
          &  LASSO  &  0.45  &  0.42  &  0.40  \\
\hline
      &  RFR     &  0.34  &  0.33  &  0.30  \\
Test  &  KRR    &  0.36  &  0.35  &  0.31  \\
      &  LASSO  &  0.43  &  0.40  &  0.41  \\
\hline
                        &  RFR     &  0.35  &  0.33  &  0.30  \\
CdTe$_{0.5}$Se$_{0.5}$  &  KRR    &  0.36  &  0.30  &  0.34  \\
                        &  LASSO  &  0.42  &  0.34  &  0.35  \\
\hline
                       &  RFR     &  0.35  &  0.34  &  0.33  \\
CdSe$_{0.5}$S$_{0.5}$  &  KRR    &  0.40  &  0.42  &  0.37  \\
                       &  LASSO  &  0.49  &  0.46  &  0.44  \\
\hline
\end{tabular}
\end{table*}

We trained regression models in a similar fashion for $\epsilon$(q/q-1) impurity transition levels. In this case, we add two additional descriptors to the earlier sets: the impurity atom oxidation state (\textit{O$_1$}) and the oxidation state (\textit{O$_2$}) of the defect site atom (+2 for Cd, -2 for Te/Se/S, and 0 for interstitial), such that \textit{O$_1$} - \textit{O$_2$} = \textit{q}; this enables the training of one model for $\epsilon$(q/q-1), rather than separate models for +2/+1, 0/-1, etc. Fig. \ref{Fig:Learning} (d) and (e) show the prediction RMSE for test and out-of-sample points respectively using the three sets of descriptors (each containing \textit{O$_1$} and \textit{O$_2$} as additional dimensions) as a function of the training set size. While the errors steadily go down with infusion of more training data, there is only a slight improvement in prediction performances going from elemental to unit cell defect properties as descriptors. The respective feature importance values (in \%, obtained from the random forest algorithm) have been listed for different RFR models in Table SI-2; it can be seen that while the unit cell defect formation enthalpy has the highest importance for predicting $\Updelta$H, as follows from Fig. \ref{Fig:Learning}(a), the impurity atom oxidation state \textit{O$_1$} shows the highest importance for $\epsilon$(q/q-1). Despite the notable correlation between certain transition levels like +1/0 and 0/-1 and the band edges from unit cell defect calculations, the improvement in prediction performance upon adding unit cell defect properties is less drastic; regardless, the best accuracies are still obtained while using the elemental + unit cell defect properties as descriptors. \\

From Fig. \ref{Fig:Learning} (d) and (e), it can be seen that the RMSE gradually saturates to around 0.31 eV for the test set and 0.34 eV for the out-of-sample points. Further, $\epsilon$(q/q-1) prediction RMSE are listed for RFR, KRR and LASSO models (using 90$\%$ of the dataset of CdTe, CdSe, and CdS points for training) in Table \ref{table:PBE_CT}; KRR predictions when using the elemental + unit cell defect properties are comparably good whereas LASSO errors are higher. Parity plots for the best RFR models trained for $\epsilon$(q/q-1) are presented in Fig. \ref{Fig:PBE_parity}, with performances shown (along with the uncertainties) for the training and test points in (c) and for the out-of-sample CdTe$_{0.5}$Se$_{0.5}$ and CdSe$_{0.5}$S$_{0.5}$ points in (d). Parity plots for models trained using KRR and LASSO are shown in Fig. SI-15. \\

We have seen that predictive models can be trained for both $\Updelta$H and $\epsilon$(q/q-1) using a set of elemental properties and unit cell defect properties as descriptors, and predictions can be made with high accuracy for impurities in out-of-sample mixed-anion compounds. With this confidence, we use the models presented in Fig. \ref{Fig:PBE_parity} to predict the impurity formation enthalpies and charge transition levels (at the PBE level of theory) respectively for all impurities in CdTe, CdSe, CdS, CdTe$_{0.5}$Se$_{0.5}$, and CdSe$_{0.5}$S$_{0.5}$. Before making these predictions for the entire chemical space and using them to screen candidates that act as `dominating' impurities, we explore the possibility of training such models for the HSE06 $\epsilon$(q$_1$/q$_2$) values. It should be noted that the PBE computed transition levels have been shown to span the physical band gap of the semiconductor \cite{Defect_DFT_6}, and also known to match well with HSE computed values \cite{HSE3}. As we discuss later, both the PBE and HSE transition levels can compare well with experimentally measured values. In the next section, we present a smaller computational dataset of impurity levels computed using the HSE06 functional, and train predictive models for the same. \\

\subsection{DFT Data and ML Models: HSE $\epsilon$(q$_1$/q$_2$)}

\begin{figure}[htp]
\includegraphics[width=\linewidth]{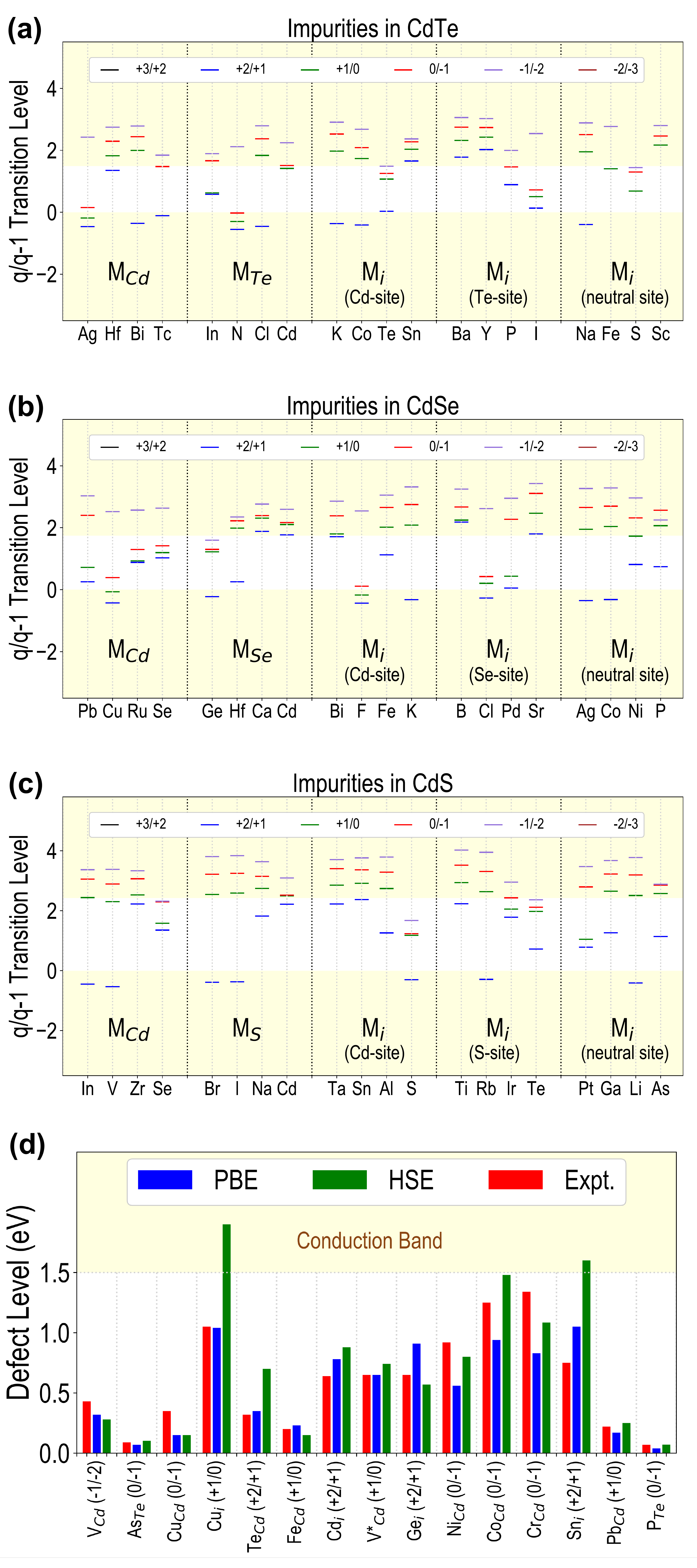}
\caption{\label{Fig:HSE_CT} Charge transition levels (from +2/+1 to -1/-2) calculated at the HSE06 level of theory for selected impurity atoms in different sites in (a) CdTe, (b) CdSe and (c) CdS. In (d), we present a comparison between experimentally measured defect levels \cite{Expt_comp1,Expt_comp2,Expt_comp3,Expt_comp4,Expt_comp5,Expt_comp6} and the corresponding PBE and HSE computed values in this work. V$_\mathrm{Cd}$ refers to a Cd vacancy whereas V*$_\mathrm{Cd}$ is the Vanadium at Cd site impurity.}
\end{figure}

For the HSE06 impurity calculations, we consider the same chemical space of 63 elements as impurity atoms, and for selected impurities, we calculated 4 transition levels (+2/+1, +1/0, 0/-1, -1/-2), since a large majority of the impurity levels that occur within the band gap or around the band edges belong to one of these 4 transitions. Because of the reliability of PBE formation enthalpies in screening low energy impurities, and the requirement of HSE-based chemical potentials of relevant species, we calculated only $\epsilon$(q$_1$/q$_2$) and not $\Updelta$H at the HSE level of theory. As shown in Table \ref{table:dataset}, we generate computational data for 19$\%$ of the total CdTe points, about 10$\%$ each of the CdSe and CdS points, and about 5$\%$ each of the points belonging to CdTe$_{0.5}$Se$_{0.5}$ and CdSe$_{0.5}$S$_{0.5}$. This totals to a dataset of less than 10$\%$ of the entire space of HSE $\epsilon$(q$_1$/q$_2$) levels in the 5 compounds. A glimpse of this dataset is provided in Fig. \ref{Fig:HSE_CT}; the +2/+1 to -1/-2 transition levels are plotted for selected impurities in the 5 defect sites in (a) CdTe, (b) CdSe, and (c) CdS. The entire HSE computational data has been plotted in Fig. SI-16 to SI-20. \\

It is interesting to note from a comparison between Fig. \ref{Fig:PBE_data} and Fig. \ref{Fig:HSE_CT} that for a given set of impurities, the observed transition levels might occur at different absolute positions but follow the same qualitative trend. For instance, going from Pb$_\mathrm{Cd}$ to Cu$_\mathrm{Cd}$ to Ru$_\mathrm{Cd}$ to Se$_\mathrm{Cd}$ in CdSe, the +2/+1 impurity level first goes down and then rises towards the CBM in both PBE and HSE. However, Ru$_\mathrm{Cd}$ and Se$_\mathrm{Cd}$ exhibit +2/+1 levels deeper in the band gap in HSE than PBE. The same trend can be seen across the PBE and HSE values of the +2/+1 and +1/0 levels for Pt$_\mathrm{i}$, Ga$_\mathrm{i}$, Li$_\mathrm{i}$ and As$_\mathrm{i}$ at the neutral interstitial site in CdS. A plot between the PBE and HSE $\epsilon$(q$_1$/q$_2$) in Fig. SI-13 shows that there is a very high correlation between the two; the HSE values lie between the y=x and the y=x+1 lines. We also collected some experimentally measured defect levels in CdTe from the literature \cite{Expt_comp1,Expt_comp2,Expt_comp3,Expt_comp4,Expt_comp5,Expt_comp6} and plotted a comparison between experiments, PBE $\epsilon$(q$_1$/q$_2$), and HSE $\epsilon$(q$_1$/q$_2$), for various defects in Fig. \ref{Fig:HSE_CT}(d). It can be seen that in general, there is good correspondence between the three, with the exception of a couple of cases where the HSE value is highly overestimated (Cu and Sn interstitial defects). Based on these 15 data points, PBE $\epsilon$(q$_1$/q$_2$) shows an RMSE of 0.22 eV with respect to experiments, whereas HSE $\epsilon$(q$_1$/q$_2$) shows a higher RMSE of 0.35 eV. There could be many reasons for this discrepancy, such as the requirement of a different mixing parameter \cite{HSE_mixing}, but is should be noted that the RMSE for HSE $\epsilon$(q$_1$/q$_2$) drops to 0.18 eV when Cu$_\mathrm{i}$ and Sn$_\mathrm{i}$ are removed. While the PBE transition levels can be assumed to be reliable, predictions at the HSE level of theory are certainly useful. \\

\begin{figure}[htp]
\includegraphics[width=\linewidth]{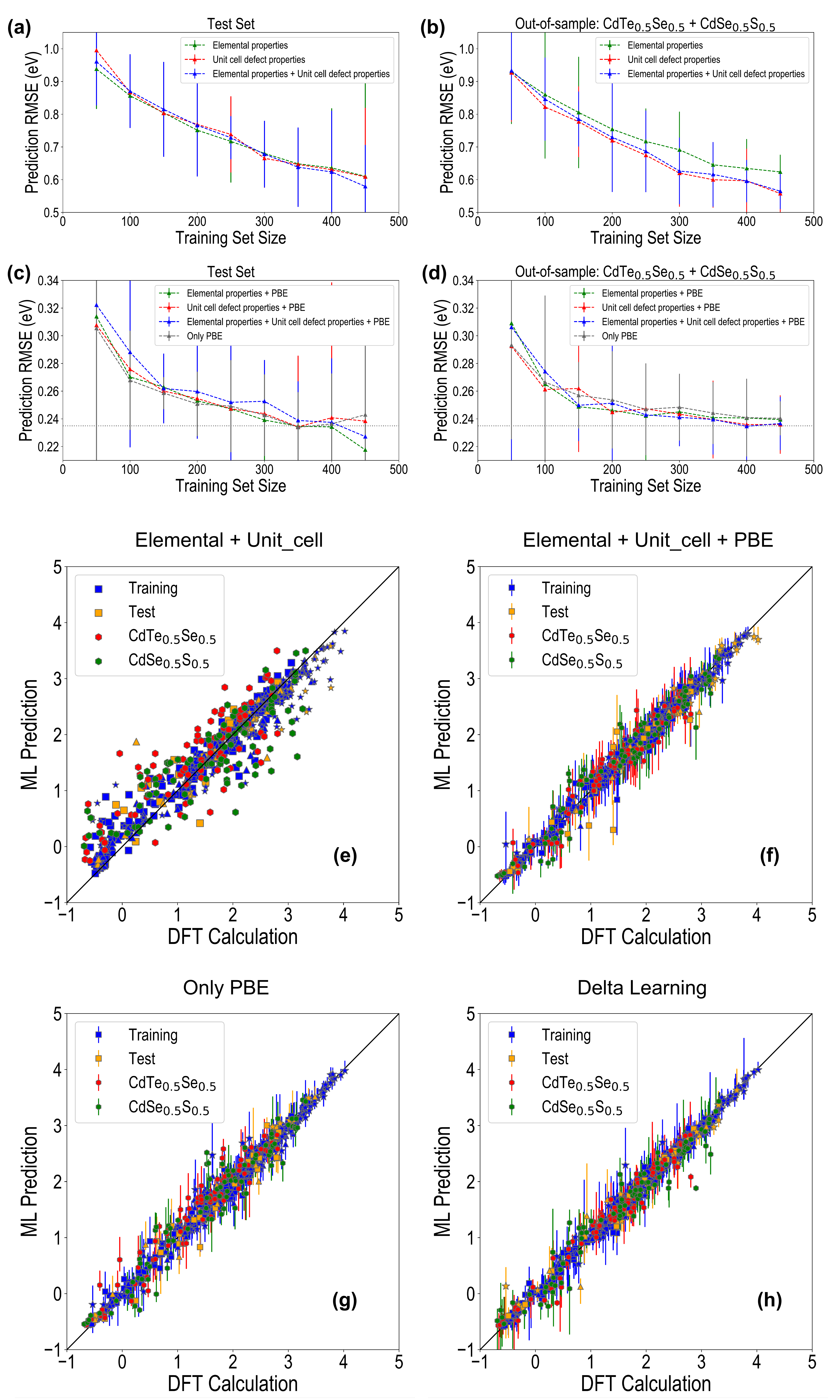}
\caption{\label{Fig:HSE_CT_learn} Prediction RMSE plotted against the training set size for random forest regression models trained for $\epsilon$(q$_1$/q$_2$) (at the HBE level of theory), using different sets of features, for (a) the test set points, without using PBE, (b) the out-of-sample points, without using PBE, (c) the test set point, using PBE as a descriptor, and (d) the out-of-sample points using PBE as a descriptor. Further, parity plots are shown for predictive models trained using 90$\%$ of the CdTe+CdSe+CdS dataset as the training set, with performances shown for the training, test and out-of-sample points, using the elemental and unit cell defect descriptors (e) without PBE and (f) with PBE, (g) using just the PBE values as descriptor, and (h) using Delta learning. Uncertainties are not plotted in (e) because they are very high in general.}
\end{figure}

We applied the same descriptors as before to train regression models for the smaller dataset of HSE transition levels, but also used the PBE $\epsilon$(q$_1$/q$_2$) as additional descriptors. Similar to Fig. \ref{Fig:Learning}(a), the linear correlation coefficient plot in Fig. SI-21 shows that while the HSE $\epsilon$(q$_1$/q$_2$) levels have high correlation with certain unit cell defect properties, the correlation between HSE and PBE $\epsilon$(q$_1$/q$_2$) is > 0.95. In Fig. \ref{Fig:HSE_CT_learn}, we plotted the prediction RMSE as a function of the training set size for the test and out-of-sample sets for RFR models trained for HSE $\epsilon$(q$_1$/q$_2$) using various combinations of descriptors. Fig. \ref{Fig:HSE_CT_learn} (a) and (b) show the errors using the usual three sets of descriptors as before; it is seen that the performances are nearly identical for the test set across the three descriptor sets, while the unit cell defect properties improve the performances for impurities in CdTe$_{0.5}$Se$_{0.5}$ and CdSe$_{0.5}$S$_{0.5}$. Error saturation is not quite seen when using more than 90$\%$ of the CdTe, CdSe, and CdS data for training, which implies that more data is potentially required for training accurate and generalizable models. \\

In Fig. \ref{Fig:HSE_CT_learn} (c) and (d), we plotted the prediction RMSE for the test and out-of-sample points respectively, using descriptor sets that include the PBE $\epsilon$(q$_1$/q$_2$) values as an added dimension. It can be seen that there is a drastic improvement in prediction performances and both test and out-of-sample errors seem to saturate around 0.24 eV. Further, we trained RFR models for HSE $\epsilon$(q$_1$/q$_2$) using only the PBE $\epsilon$(q$_1$/q$_2$) value as sole descriptor, and see that predictions are similar to the other three sets of descriptors. In Fig. \ref{Fig:HSE_CT_learn} (e) to (h), we present four different predictive models; (e), (f), and (g) show RFR models trained using different sets of descriptors, and it can be seen that the addition of PBE values as descriptors significantly improves the performance. This can also be seen from the RMSE values listed in Table \ref{table:HSE_CT}--including PBE $\epsilon$(q$_1$/q$_2$) as a descriptor brings down the test and out-of-sample RMSE to $\sim$ 0.20 eV. We further applied a technique called Delta-learning, wherein we train RFR models for the difference between HSE and PBE transition levels ($\delta$ property = HSE $\epsilon$(q$_1$/q$_2$) - PBE $\epsilon$(q$_1$/q$_2$)), and predict HSE $\epsilon$(q$_1$/q$_2$) values by adding the predicted $\delta$ property to PBE $\epsilon$(q$_1$/q$_2$). It can be seen from Fig. \ref{Fig:HSE_CT_learn} (h) that very low test (RMSE = 0.21 eV) and out-of-sample (RMSE = 0.22 eV) errors can be obtained for Delta-learning. Overall, it is seen that the RFR model trained for HSE $\epsilon$(q$_1$/q$_2$) using the elemental properties + unit cell defect properties + PBE $\epsilon$(q$_1$/q$_2$) as descriptors gives the lowest test set and out-of-sample errors, and can be used for making predictions for the > 90$\%$ of the dataset yet to be computed. Predictive models trained for HSE $\epsilon$(q$_1$/q$_2$) using KRR and LASSO are presented in Fig. SI-22 for comparison with RFR. \\

\begin{table*}[htp]
\centering
\caption{RMSE (in eV) for regression models trained for HSE $\epsilon$(q$_1$/q$_2$), using different methods and sets of features.}
\label{table:HSE_CT}
\medskip
\begin{tabular}{c|c|c|c|c|c|c|c}
\hline
\textbf{Dataset} & \textbf{Regression Method} & \multicolumn{2}{c|}{\textbf{Elemental properties}} & \multicolumn{2}{c|}{\textbf{Unit cell defect properties}} & \multicolumn{2}{c}{\textbf{Elemental + Unit cell defect properties}}  \\
\hline
          &       &  \textbf{Without PBE}  &  \textbf{With PBE}  &  \textbf{Without PBE}  &  \textbf{With PBE}  &  \textbf{Without PBE}  &  \textbf{With PBE}  \\
\hline
          &  RFR     &  0.31  &  0.10  &  0.28  &  0.10  &  0.28  &  0.10  \\
Training  &  KRR    &  0.47  &  0.29  &  0.40  &  0.28  &  0.48  &  0.28  \\
          &  LASSO  &  0.70  &  0.22  &  0.63  &  0.22  &  0.60  &  0.21  \\
          &  $\delta$-learn (RFR)  &  0.14  &  0.10  &  0.13  &  0.10  &  0.14  &  0.09  \\
\hline
          &  RFR     &  0.61  &  0.23  &  0.63  &  0.24  &  0.62  &  0.24  \\
Test      &  KRR    &  0.61  &  0.28  &  0.57  &  0.29  &  0.58  &  0.30  \\
          &  LASSO  &  0.72  &  0.24  &  0.65  &  0.24  &  0.63  &  0.24  \\
          &  $\delta$-learn (RFR)  &  0.28  &  0.20  &  0.28  &  0.22  &  0.27  &  0.21  \\
\hline
                        &  RFR     &  0.64  &  0.22  &  0.59  &  0.21  &  0.58  &  0.21  \\
CdTe$_{0.5}$Se$_{0.5}$  &  KRR    &  0.60  &  0.52  &  0.54  &  0.53  &  0.55  &  0.53  \\
                        &  LASSO  &  0.71  &  0.17  &  0.54  &  0.17  &  0.55  &  0.16  \\
                        &  $\delta$-learn (RFR)  &  0.25  &  0.20  &  0.20  &  0.19  &  0.21  &  0.18  \\
\hline
                       &  RFR     &  0.63  &  0.27  &  0.61  &  0.26  &  0.61  &  0.26  \\
CdSe$_{0.5}$S$_{0.5}$  &  KRR    &  0.57  &  0.50  &  0.62  &  0.51  &  0.52  &  0.50  \\
                       &  LASSO  &  0.71  &  0.23  &  0.60  &  0.22  &  0.61  &  0.22  \\
                       &  $\delta$-learn (RFR)  &  0.28  &  0.26  &  0.26  &  0.25  &  0.26  &  0.25  \\
\hline
\end{tabular}
\end{table*}

\subsection{Screening of impurities for Fermi level tuning}

Using predictions for the neutral state impurity formation enthalpy $\Updelta$H, and every impurity transition level $\epsilon$(q$_1$/q$_2$) from +3/+2 to -2/-3, the Fermi level (E$_F$) and charge (q) dependent formation enthalpy (E$^f$) can be predicted for every possible impurity in Cd-rich or anion-rich chemical potential conditions. For this analysis, we use the machine-learned predictions at the PBE level of theory, since that the formation enthalpies are known to be qualitatively reliable and the transition levels match well with reported experiments, as shown in Fig. \ref{Fig:HSE_CT}(d). In the absence of any external impurities, the equilibrium Fermi level in a semiconductor is determined by its dominant native point defects, such as vacancies or self-interstitial defects. By comparing the machine-learned formation enthalpy of any impurity with the computed energetics of dominant intrinsic defects, we can estimate the probable change in the nature of conductivity that would occur upon introduction of the impurity in the semiconductor. In order to go through this process, we simulated all possible vacancy (e.g., V$_\mathrm{Cd}$, which refers to a Cd vacancy), self-interstitial (e.g., Cd$_\mathrm{i}$ or Se$_\mathrm{i}$) and anti-site defects (e.g., Cd$_\mathrm{Te}$, S$_\mathrm{Cd}$, etc.) in supercells of CdTe, CdSe, CdS, CdTe$_{0.5}$Se$_{0.5}$, and CdSe$_{0.5}$S$_{0.5}$. The DFT computed E$^f$ vs E$_F$ plots for all possible intrinsic defects in the 5 compounds are presented in Figures SI-23 to SI-27. \\

The computed energetics of intrinsic defects reveal that while the Cd vacancy, V$_\mathrm{Cd}$, is the dominant acceptor type defect in each compound, the Cd interstitial defect, Cd$_\mathrm{i}$ \textit{(Te-site)}, is the dominant donor type defect in CdTe, CdSe, and CdS, and Cd interstitial defect, Cd$_\mathrm{i}$ \textit{(Cd-site)}, is the dominant donor type defect in CdTe$_{0.5}$Se$_{0.5}$ and CdSe$_{0.5}$S$_{0.5}$. It is also seen that the equilibrium Fermi level (determined using charge neutrality conditions \cite{fermi_eqm}) is near the middle of the band gap for Cd-rich conditions in every compound, which would lead to an intrinsic type of conductivity. The equilibrium E$_{F}$ shifts towards the valence band upon going from Cd-rich to anion-rich conditions, and in all cases renders the conductivity moderately p-type. If an impurity creates a charged defect that is more stable within the band gap than either the dominant acceptor or donor type defect, it can pin the Fermi level at a different location and change the conductivity. We predicted the E$^f$ values of every possible impurity in the 5 compounds as a function of E$_F$, and screened those impurities which would cause a shift in the equilibrium E$_F$. The complete list of all such `dominating impurities' is provided in Tables SI-1 to SI-5. The dominating defects under Cd-rich and Te-rich conditions and the nature of conductivity are in agreement with reported literature \cite{Intrinsic}. \\

\begin{figure*}[t]
\includegraphics[width=\linewidth]{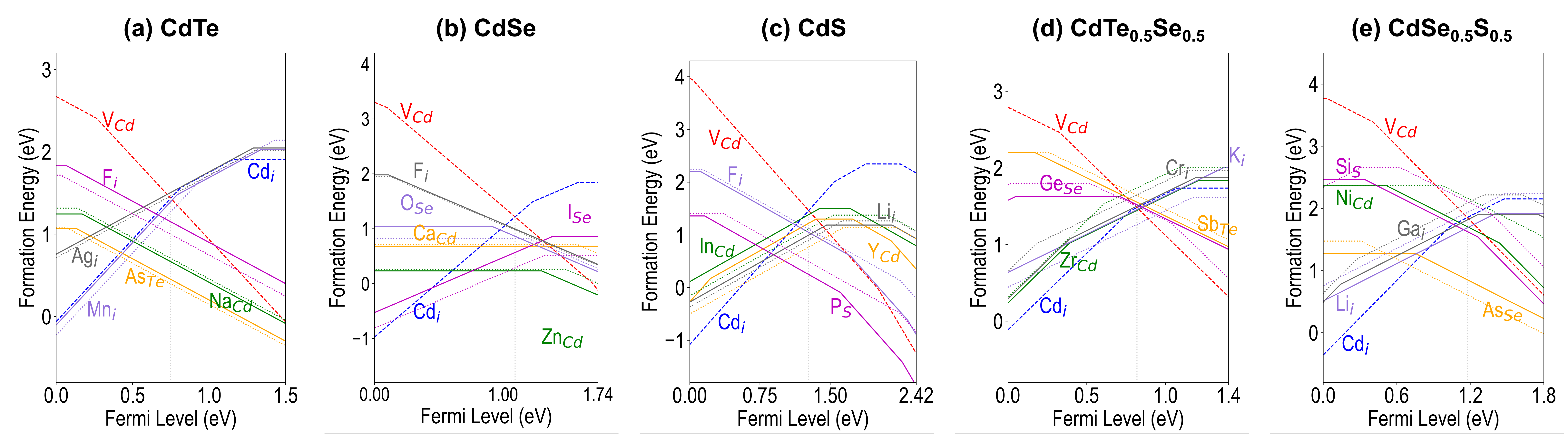}
\caption{\label{Fig:Defect_form_1} Machine learned defect formation energies at Cd-rich chemical potential conditions for selected impurities predicted to shift the equilibrium Fermi level in (a) CdTe, (b) CdSe, (c) CdS, (d) CdTe$_{0.5}$Se$_{0.5}$, and (e) CdSe$_{0.5}$S$_{0.5}$. The intrinsic defects are shown as dashed lines and ML predictions of different impurities as solid lines, while the dotted lines represent the computed formation energies from DFT; it can be seen that DFT and ML match pretty well.}
\end{figure*}

Given that both DFT and ML predictions of $\Updelta$H and $\epsilon$(q$_1$/q$_2$) are available for all 315 possible impurity-site combinations in CdTe, we compare the E$^f$ vs E$_F$ plots for each impurity estimated from both methods. Specifically, the evaluation of an impurity as shifting or not shifting the equilibrium E$_F$ (alternatively, whether the impurity dominates over the intrinsic defects or not) is used as a metric to compare the DFT and ML predictions. We present such a comparison in Table \ref{table:Verdict} in terms of the total number of false and true positives or negatives predicted by ML for impurities in Cd-rich and Te-rich chemical potential conditions. It is seen that the false negatives and false positives amount to less than 5$\%$ of the total impurities, which means that the ML approach has a > 95$\%$ probability of successful classification of an impurity as dominating or not. The true positives, which are the impurities predicted to be dominating by both DFT and ML, amount to about 30 in total for both Cd-rich and Te-rich conditions. The total number of dominating impurities as predicted by ML for the 5 compounds (and listed in Tables SI-3 to SI-7) in Cd-rich and anion-rich conditions are presented in Table \ref{table:Dominating}. \\

For a few selected `dominating' impurities, we plotted the ML predicted E$^f$ as a function of E$_F$ in Fig. \ref{Fig:Defect_form_1} for (a) CdTe, (b) CdSe, (c) CdS, (d) CdTe$_{0.5}$Se$_{0.5}$, and (e) CdSe$_{0.5}$S$_{0.5}$, for Cd-rich chemical potential conditions. The formation energies of V$_\mathrm{Cd}$ and Cd$_\mathrm{i}$ are plotted (using dashed lines) as well to illustrate how each impurity dominates and changes the equilibrium E$_F$. Additional DFT computations (wherever missing) were performed for these selected dominating impurities; the DFT computed E$^f$ is plotted in each case using dotted lines, and it can be seen that there is a very good match between the DFT and ML predicted lines. Impurities such as Na$_\mathrm{Cd}$, Zn$_\mathrm{Cd}$, F$_\mathrm{i}$ and Cu$_\mathrm{Cd}$ create acceptor type defects, whereas impurities like Mn$_\mathrm{i}$, Bi$_\mathrm{Cd}$, Cl$_\mathrm{Se}$ and Li$_\mathrm{i}$ are donor type. A common thread across the 5 compounds is low energy defects created by Group I elements and certain transition metals at the Cd-site, halogen atoms and Group V atoms at the X-site, and F, Li and Ag at the interstitial sites. Indeed, there is abundant experimental literature on using a variety of dopants to change the properties of CdTe, such as p-type doping using As$_\mathrm{Te}$ \cite{Doping_expt1}, Sb$_\mathrm{Te}$ \cite{Doping_expt2} and Na$_\mathrm{Cd}$ \cite{Doping_expt3}, and improved solar cell efficiency using halogen atoms \cite{Doping_expt4}, Zn$_\mathrm{Cd}$ doping \cite{Doping_expt5}, and Li$_\mathrm{Cd}$ or Li$_\mathrm{i}$ \cite{Doping_expt6}. In summary, ML has successfully screened all the impurities that can potentially be introduced in these Cd-chalcogenides to alter the conductivity type and consequently the semiconductor's optoelectronic properties. \\

\begin{table}[htp]
\centering
\caption{A comparison between predictions by DFT and ML of `dominating impurities' in CdTe. True positives refer to the cases that were predicted to be dominating by both DFT and ML, and true negatives are the cases predicted to be non-dominating by both. False positives were predicted to be dominating by only ML whereas false negatives were predicted to be dominating by only DFT.}
\label{table:Verdict}
\medskip
\begin{tabular}{c|c|c|c|c}
\hline
\textbf{Verdict} & \multicolumn{2}{c|}{\textbf{Cd-rich}} & \multicolumn{2}{c}{\textbf{Te-rich}} \\
\hline
  &  \textbf{Predicted}  &  \textbf{$\%$ of Total}  &  \textbf{Predicted}  &  \textbf{$\%$ of Total} \\
\hline
 False Positives     &  5  &  1.59  &  3  &  0.95  \\
 False Negatives     &  10  &  3.17  &  6  &  1.90  \\
 True Negatives     &  272  &  86.35  &  275  &  87.30  \\
 True Positives     &  28  &  8.89  &  31  &  9.84  \\
\hline
\end{tabular}
\end{table}

\begin{table}[htp]
\centering
\caption{The total number of impurities predicted to be dominating by ML for Cd-rich and anion-rich chemical potential conditions in the 5 CdX compounds.}
\label{table:Dominating}
\medskip
\begin{tabular}{c|c|c|c|c}
\hline
\textbf{CdX} & \multicolumn{2}{c|}{\textbf{Cd-rich}} & \multicolumn{2}{c}{\textbf{Te-rich}} \\
\hline
  &  \textbf{Predicted}  &  \textbf{$\%$ of Total}  &  \textbf{Predicted}  &  \textbf{$\%$ of Total} \\
\hline
 CdTe     &  28 / 315  &  8.89  &  31 / 315  &  9.84  \\
 CdSe     &  24 / 315  &  7.62  &  18 / 315  &  5.71  \\
 CdS     &  15 / 315  &  4.76  &  21 / 315  &  6.67  \\
 CdTe$_{0.5}$Se$_{0.5}$ &  44 / 441  &  9.98  &  31 / 441  &  7.03  \\
 CdSe$_{0.5}$S$_{0.5}$ &  36 / 441  &  8.16  &  26 / 441  &  5.90  \\
\hline
\end{tabular}
\end{table}

\section*{Conclusions and Outlook}

In this work, we showed that machine learning can be used to train accurate predictive models of the formation enthalpy ($\Updelta$H) and defect transition levels ($\epsilon$(q$_1$/q$_2$)) of impurities in Cd-based chalcogenides using DFT generated data. The choice of descriptors is of vital importance; we see that combining elemental properties of an impurity atom with energetic and electronic information computed from a lower-cost unit cell defect calculation leads to the optimal set of features that serve as inputs to random forest regression models. Predictive models thus trained for $\Updelta$H and $\epsilon$(q$_1$/q$_2$) using data generated for CdTe, CdSe, and CdS at the PBE level of theory can accurately predict the impurity properties of mixed anion compounds CdTe$_{0.5}$Se$_{0.5}$ and CdSe$_{0.5}$S$_{0.5}$, showing their true out-of-sample predictive power. Models were further trained and tested for a smaller dataset of $\epsilon$(q$_1$/q$_2$) values at the HSE level of theory, for which the use of PBE $\epsilon$(q$_1$/q$_2$) as a descriptor leads to significant improvement in prediction performances. The trained models were used to make predictions for the entire chemical space of impurities in the 5 compounds, following which the formation enthalpy (E$^f$) of every impurity was obtained as a function of the Fermi level (E$_F$) in the band gap. The E$^f$ vs E$_F$ behavior is used to determine whether an impurity can shift the equilibrium E$_F$ in the semiconductor as determined by the dominant intrinsic point defects, leading to a list of impurities in each compound that can dominate over the intrinsic defects and change the nature of conductivity in the material. A comparison of DFT and ML predictions shows that less than 5$\%$ of the entire population of impurities in CdTe is classified as false negative or positive (in terms of its `dominating' nature), giving us confidence that this ML approach can be used for a successful screening of stable and active impurity atoms in preferred defect sites. \\

The combined DFT and ML approach demonstrated here can be applied to any number of semiconductor classes. For instance, III-V semiconductors such as GaN, GaP, GaAlP, AlP, BP etc. are interesting materials for photodiodes, solar cells, and in recent times, have been studied for intermediate band photovoltaic applications \cite{IB9,IB10,IB11,IB12}. A quick screening of impurity atoms that can not only change the equilibrium Fermi level, but also create energy level(s) in the band gap, can be made possible using machine learned models to predict impurity properties. Given the ubiquity of the descriptors used here, this approach can, in theory, be extended to include all possible pure and mixed compositions of II-VI, III-V and group IV semiconductors, many of which are currently serving various optoelectronic applications. Further extensions can be made in terms of impurity atoms by including the lanthanides and actinides as well. There are also opportunities in applying a wide variety of descriptors for further improvement in ML performance, such as using Coulomb matrix representation, radial distribution function, or electron density distribution. A true `semiconductor+impurity' design framework will be complete once the forward prediction model is combined with an inverse model as well, wherein genetic algorithms or other optimization techniques are used to devise suitable compositions which lead to stable impurities with favorable energy levels in the band gap. \\


\section*{METHODS}

\subsection*{DFT Details}

We used 2$\times$2$\times$2 supercells for any CdX compound, resulting in a system with 64 atoms, to optimize the (fixed cell shape and size) geometry using DFT in the neutral and charged states. The starting structures of CdTe, CdSe, and CdS were obtained from the Materials Project \cite{MP}. The structures of CdTe$_{0.5}$Se$_{0.5}$ and CdSe$_{0.5}$S$_{0.5}$ were simulated using special quasi-random structures \cite{QSR} generated from CdTe and CdSe respectively. The computed lattice constants of the 5 compounds are listed in Table SI-1. DFT computations were performed using the Vienna ab-initio Simulation Package (VASP) employing the Perdew-Burke-Ernzerhof (PBE) exchange-correlation functional and projector-augmented wave (PAW) atom potentials. The kinetic energy cut-off for the planewave basis set was 400 eV, and all atoms were relaxed until forces on each were less than 0.05 eV/{\text{\AA}}. Brillouin zone integration was performed using a 3$\times$3$\times$3 Monkhorst-Pack mesh. Further, HSE06 calculations were performed for a smaller dataset using a 4$\times$4$\times$4 Monkhorst-Pack mesh. The following equations are used to compute the formation enthalpy E$^f$ of an impurity as a function of the chemical potential $\mu$ and Fermi level E$_{F}$, and any impurity transition level, $\epsilon$(q$_1$/q$_2$) :




\begin{equation}\label{eqn-1}
\begin{multlined}
{E^{f}}(q,E_{F}) = E(D^{q}) - E(CdX) + {\mu} + q(E_{F} + E_{vbm}) + E_{corr}
\end{multlined}
\end{equation} 
\begin{equation}\label{eqn-2}
\begin{multlined}
{\epsilon}(q_1/q_2) = \frac{E^{f}(q_1) - E^{f}(q_2)}{q_{2}-q_{1}}
\end{multlined}
\end{equation}

E(\textit{D$^{q}$}) and E(\textit{CdX}) refer to the total DFT energy of the defect containing system in charge \textit{q} and the bulk CdX compound, respectively. E$_{vbm}$ refers to the valence band maximum of bulk CdX and E$_{corr}$ is the correction energy necessary due to periodic interaction between charges \cite{Corr1,Corr2}.

\subsection*{Regression Techniques}

RFR is based on ensemble learning through decision trees, where each tree is built using bootstrap samples randomly drawn from the dataset. By optimizing the number of trees and the number of necessary features, RFR prepares a final predictive model as an ensemble, provides errors bars in predictions based on standard deviation across individual trees, and assigns a relative importance to the different features. KRR is a similarity based regression algorithm where the output is expressed as a weighted sum over Kernel functions, which are defined in terms of the Euclidean distance between data points (which is a measure of the similarity). We use a Gaussian kernel in this work, and the hyperparameters that are optimized are the Kernel coefficients and the Gaussian width. LASSO is similar to ridge regression but uses an L1 regularization, unlike KRR which uses L2 regularization. LASSO regression operates on the principle of shrinking the coefficients of many features down to zero, and is thus very useful when there are a large number of features. More details about random forest regression, Kernel ridge regression, and LASSO regression can be obtained from references 32, 34, and 35, respectively. Each technique was applied on the DFT data using python packages available in Scikit-learn (https://scikit-learn.org/stable/).


\section*{ACKNOWLEDGMENTS}


We acknowledge funding from the US Department of Energy SunShot program under contract DOE DEEE005956. Use of the Center for Nanoscale Materials, an Office of Science user facility, was supported by the U.S. Department of Energy, Office of Science, Office of Basic Energy Sciences, under Contract No. DE-AC02-06CH11357. We gratefully acknowledge the computing resources provided on Bebop, a high-performance computing cluster operated by the Laboratory Computing Resource Center at Argonne National Laboratory. This research used resources of the National Energy Research Scientific Computing Center, a DOE Office of Science User Facility supported by the Office of Science of the U.S. Department of Energy under Contract No. DE-AC02-05CH11231. MYT would like to acknowledge support from the U.S. Department of Energy, Office of Science, Office of Workforce Development for Teachers and Scientists (WDTS) under the Science Undergraduate Laboratory Internship (SULI) program. MJD was was supported by the U. S. Department of Energy , Office of Basic Energy Sciences, Division of Chemical Sciences, Geosciences, and Biosciences, under Contract No. DE-AC02-06CH11357.

\subsection*{Author Contributions} M.K.Y.C., R.F.K. and A.M.K. conceived the idea. A.M.K., M.Y.T. and F.G.S. performed the DFT computations. A.M.K. and M.J.D. trained ML models. All authors contributed to the discussion and writing of the manuscript.

\subsection*{Data Availability} All the ab-initio data and ML models will be made available to the public upon publication of the manuscript.

\subsection*{Additional Information}

The authors declare no competing financial or non-financial interests. \\
 
Correspondence and requests for materials should be addressed to A.M.K. (email:mannodiarun@anl.gov) or M.K.Y.C. (email: mchan@anl.gov).


\section*{REFERENCES}

\bibliographystyle{naturemag}
\bibliography{mybibfile}

\end{document}